\documentclass[10pt]{article}

\usepackage{graphicx,epsf,color,tabularx}
\usepackage{amsfonts,amsthm,amsmath,bbm}

\def\D{{\cal D}}
\def\F{{\cal F}}
\def\H{{\cal H}}
\def\I{{\cal I}}
\def\K{{\cal K}}

\def\B{{\cal B}}
\def\Rset{\mathbb{R}}
\def\Sset{\mathbb{S}}
\def\opt{\mathrm{opt}}
\def\diag{\mathrm{diag}}
\def\sign{\mathrm{sign}}
\renewcommand\vec[1]{\boldsymbol{#1}}

\newtheorem{proposition}{Proposition}
\newtheorem{corollary}{Corollary}

\title{On a generalized entropic uncertainty relation in the case of the qubit}

\author{S.\ Zozor $^{1,2}$, G.\ M.\ Bosyk $^{2}$, and M.\ Portesi $^{2}$
\\
\footnotesize $^1$~{Laboratoire Grenoblois d'Image, Parole, Signal et Automatique (GIPSA-Lab, CNRS), 961 rue de la Houille Blanche,} \\
\footnotesize {38402 Saint Martin d'H\`eres, France}
\\
\footnotesize $^2$~{Instituto de F\'{\i}sica La Plata (IFLP), CONICET, and Departamento de F\'{\i}sica, Facultad de Ciencias Exactas,} \\
\footnotesize {Universidad Nacional de La Plata, C.C.~67, 1900 La Plata, Argentina}
}

\date{\today}

\begin{document}

\maketitle


\begin{abstract}

  We revisit generalized entropic  formulations of the uncertainty principle for
  an  arbitrary pair of quantum observables in  two-dimensional Hilbert  space. R\'enyi
  entropy  is  used as  uncertainty  measure  associated  with the  distribution
  probabilities corresponding  to the outcomes  of the observables. We  derive a
  general expression for  the tight lower bound of the  sum of R\'enyi entropies
  for any couple of (positive)  entropic indices $(\alpha,\beta)$. Thus, we have
  overcome the H\"older conjugacy constraint  imposed on the entropic indices by
  Riesz--Thorin theorem.   In addition, we present an  analytical expression for
  the   tight  bound   inside  the   square  $\left[   0  \:   ,  \:   \frac  12
  \right]^2$
  in the  $\alpha$--$\beta$
  plane, and  a semi-analytical expression on  the line $\beta=  \alpha$.  It is
  seen  that previous  results are  included as  particular cases.  Moreover, we
  present an analytical  but suboptimal bound for any couple  of indices. In all
  cases, we provide the minimizing states.

\end{abstract}

\section{Introduction}

The uncertainty principle  (UP) is a fundamental concept  in physics that states
the  impossibility to  predict with  absolute certainty  and  simultaneously the
outcomes of measurements for pairs  of noncommuting quantum observables.  In its primary
quantitative  formulation, the  principle is  described  by the  existence of  a
nontrivial   lower   bound  for   the   product   of   the  variances   of   the
operators~\cite{Hei27,Ken27,Rob29}.  However,  such formulations are  not always
adequate  due to  various reasons.  As an  example, there  exist  variables with
infinite variance~\cite{SamTaq94}, so that the second-order moment is not always
convenient for describing the dispersion of a random variable.  Moreover, in the
case of  discrete-spectrum observables, there  is no universal  nontrivial lower
bound,   and   thus   Heisenberg-like   inequalities   do   not   quantify   the
UP~\cite{Lui01,Lui11,Zoz12}.

In order to  overcome the potential inadequacy of  the variance-based expression
of the  UP, many formulations  based on other  measures of dispersion  have been
proposed,        for        instance        issued       from        information
theory~\cite{Sha48,Ren61,CovTho06}.       The      pioneering      works      of
Hirschman~\cite{Hir57},  Bialynicki-Birula   and  Mycielski~\cite{BiaMyc75},  or
Maassen and  Uffink~\cite{MaaUff88}, have given  rise to many versions  based on
generalized          information         entropies          (or         entropic
moments)~\cite{Deu83,Bia84,Raj95,PorPla96,Bia06,ZozVig07,ZozPor08,WehWin10,Sen11,DehLop10:12},
on  Fisher  information~\cite{RomAng99,RomSan06,SanGon06},   or  on  moments  of
various orders~\cite{ZozPor11}.  Recently,  generalized versions of entropic and
support inequalities in the context  of variables described by frames instead of
bases, have been proposed~\cite{RicTor13}.

In this paper, we focus on the  R\'enyi-entropy formulation of UP in the case of
discrete-spectrum operators.   Specifically, we search for  (tight) lower bounds
for  the sum  of  R\'enyi entropies  associated with  the  outcomes of  a pair  of
observables. In the  majority of previous related studies,  the entropic indices
corresponding to both  observables are considered to be  conjugated in the sense
of  H\"older, since  the proofs  make use  of Riesz--Thorin  or Young--Hausdorff
theorems. Extensions  for nonconjugated indices  exist, based on  the decreasing
property  of  R\'enyi entropy  versus  its  index,  leading then  to  suboptimal
bounds~\cite{ZozPor08,RicTor13}. These  bounds have been refined in  the case of
2-level  systems (or  qubits) when  the entropic  indices coincide and have the value
$\frac  12$~\cite{Ras12} or  2~\cite{BosPor12,BosPor13}.  Here  we  extend these
results beyond the scope of Riesz' theorem, allowing for {\it arbitrary} couples
of indices.  We  provide a semi-analytical treatment of the  problem and we find
significant,  nontrivial inequalities  expressing  UP for  qubits. Moreover,  we
supply the minimizing states for the uncertainty relations established.

The paper  is organized  as follows. In  Sec.~\ref{Position:sec}, we  begin with
basic  definitions   and  notation,  and  summarize   known  results  concerning
generalized   entropic  uncertainty   relations  for   $N$-level   systems.   In
Sec.~\ref{Main:sec}  we state  the  problem  for qubits  and  present our  major
results. A  discussion is provided  in Sec.~\ref{Discussion:sec}. The  proofs of
our results are given in the appendices.


\section{Statement of the problem: notation and previous results}
\label{Position:sec}

We consider pairs  of quantum observables, say $A$ and $B$, with  discrete spectra on an
$N$-dimensional Hilbert space  $\H$.  Pure states $\left|\Psi\right\rangle\in\H$
can  be expanded  onto the  corresponding orthonormal  eigenbases $\{|a_k\rangle
\}_{k=1}^N$ and $\{  |b_l\rangle \}_{l=1}^N$.  In order to  fix the notation, we
write $\left| \Psi \right\rangle \: = \: \sum_{k=1}^{N} \psi_k \, |a_k\rangle \:
=  \: \sum_{l=1}^{N}  \widetilde{\psi}_l\, |b_l\rangle$ \ where the  $\psi_k$ and
$\widetilde\psi_l$  are  complex coefficients, that  we  arrange in  column
vectors:  \   $\psi  =   \left[  \psi_1  \cdots   \psi_N  \right]^t$  \   and  \
$\widetilde{\psi}   =   \left[   \widetilde{\psi}_1  \cdots   \widetilde{\psi}_N
\right]^t$ .  From orthonormality of the bases, one has
\begin{equation}
\widetilde{\psi} = T \psi \qquad \mbox{ where }
\quad T_{lk} = \langle b_l | a_k \rangle ,
\label{Relation_psi_psitilde:eq}
\end{equation}
being $T$ an $N\times N$ unitary matrix.

Vectors $\psi$  and $\widetilde{\psi}$  are such that  $\| \psi \|_2^2  = \sum_k
|\psi_k|^2 = 1$  and similarly $\| \widetilde{\psi} \|_2^2 =  1$.  A vector with
components  $|\psi_k|^2 =  \left| \langle  a_k |  \Psi \rangle  \right|^2$  is interpreted as a
probability vector,  where $|\psi_k|^2$ represents the  probability of measuring
eigenvalue $a_k$ as  outcome for observable $A$ when the quantum system  is in the state
$|\Psi\rangle$ \big(resp.\ ${|\widetilde{\psi}_l|}^2 = \left| \langle b_l | \Psi
  \rangle \right|^2$ for measurement of $B$ \big).


\hfill

We are interested in  uncertainty relations concerning simultaneous observations
of    two    magnitudes, particularly their statement   through    the    use    of    information-theoretic
quantities~\cite{CovTho06}. The measure of ignorance or lack of information that
we employ is the R\'enyi entropy of a  probability set $p = \left\{p_k : p_k  
\ge 0 , \
  \sum_{k=1}^N p_k = 1\right\}$, defined as~\cite{Ren61}
\begin{equation}
H_\lambda(p) = \frac{1}{1-\lambda} \log \left( \sum_k \, p_k^{\, \lambda} \right)
\label{Renyi:eq}
\end{equation}
where $\lambda \ge  0$ is the entropic index and $\log$  stands for natural 
logarithm. 
The limiting case $\lambda \to 1$ is well defined and gives Shannon entropy
$H_1(p) \equiv H(p) = - \sum_k p_k \log p_k$.
The  index $\lambda$  plays the  role of  a magnifying  glass, in  the following
sense:  when  $\lambda <1$,  the  contribution of  the  different  terms in  the
sum~\eqref{Renyi:eq} becomes more uniform  with respect to the case $\lambda=1$;
conversely, when  $\lambda > 1$,  the leading probabilities of  the distribution
are stressed in  the summation.  Indeed, in the  extreme case $\lambda=0$, $H_0$
is simply  a measure of the cardinality  of the support of  the probability set,
regardless of the values of the (nonzero) probabilities; this measure is closely
linked    to    the    $L^0$     norm    which    is    relevant    in    signal
processing~\cite{RicTor13,ElaBru02,GhoJam11}.   At the  opposite, $H_\infty  = -
\log(\max_k p_k)$  takes only into account  the maximum probability  of the set,
and is known  as min-entropy due to the  nonincreasing property of $H_{\lambda}$
versus  $\lambda$  for  a  given  probability  distribution.   Another  relevant
property is that R\'enyi entropy $H_\lambda$ is concave for $\lambda \in [0 \: ,
1]$, or  even when  $\lambda \in [0  \: ,  \lambda_*(N)]$ where the  upper limit
depends  on  the cardinality  of  the  probability  set [e.g.\  $\lambda_*(2)  =
2$]~\cite[p.~57]{BenZyc06}.   R\'enyi  entropies  appear  naturally  in  several
contexts, as signal  processing (Chernoff bound, Panter--Dite formula)~\cite[and
refs.\          therein]{CovTho06,Ber09,BarFla01,HerMa02},          multifractal
analysis~\cite{Har01,HenPro83,JizAri04}, or  quantum physics (collision entropy,
purity, informational energy, Gini--Simpson  index, index of coincidence, repeat
rate)~\cite[and  refs.\ therein]{DehLop10:12,JizAri04,Bas04,Jiz03,ParBir05}; see
also~\cite{Sen11} for a recent survey.

One  can easily  verify that  R\'enyi  entropies are  positive and  that in  the
$N$-states case they are upper-bounded by $\log N$:
$0 \le H_\lambda(p) \le \log N $.
The  lower   bound  is   achieved  when  the   probability  distribution   is  a
Kronecker-delta,  $p_k =  \delta_{k,i}$  for  certain $i$,  and  the upper  bound
corresponds to the uniform distribution, $p_k = 1/N$.

\hfill

In this contribution  we will consider the R\'enyi  entropies of the probability
sets   $|\psi|^2  \equiv   \{|\psi_l|^2\}$   and  $|\widetilde{\psi}|^2   \equiv
\{|\widetilde\psi_k|^2\}$,  associated with the  measurement of  observables $A$
and $B$ respectively.  Our goal is to find uncertainty relations of the type
\begin{equation}
H_\alpha(|\psi|^2) + H_\beta(|\widetilde{\psi}|^2) \ge \overline{\B}_{\alpha,\beta; N}
\end{equation}
for any couple of (positive)  entropic indices $(\alpha,\beta)$, where the bound
$\overline{\B}_{\alpha,\beta;  N}$  should  be  nontrivial,  i.e.\  nonzero,  and
universal  in  the sense of being independent  of  the  state $\left|  \Psi
\right\rangle$ of the quantum system. By definition, the tightest  bound is obtained by
minimization of the left-hand side, thus
\begin{equation}
\overline{\B}_{\alpha,\beta; N} \geq \B_{\alpha,\beta; N} \equiv
\min_{|\Psi\rangle} \: \left( H_\alpha(|\psi|^2) + H_\beta(|\widetilde{\psi}|^2)
\right)
\label{Bound_AsAMin:eq}
\end{equation}
It comes  out that  the tight  bound $\B_{\alpha,\beta; N}$  only depends  on the
transformation matrix $T$ in which  an important characteristic is the so-called
overlap (or coherence) between the eigenbases, given by
$$
c = \max_{k,l} | \langle b_l | a_k \rangle | . 
$$
From  the  unitarity  property of  matrix  $T$,  the  overlap  is in  the  range
$c\in\left[ \frac{1}{\sqrt{N}}  \: , \:  1 \right]$. The  case $c=\frac{1}{\sqrt
  N}$ corresponds  to observables $A$ and  $B$ being complementary meaning 
that maximum certainty  in the measure of one of  them implies maximum ignorance
about the other, while $c=1$ corresponds to a pair of commuting observables.


The problem has been addressed in  various contexts, and in some cases numerical
and/or  analytical  bounds  have  been  found.  Several  results  correspond  to
conjugated indices (in the sense of H\"older\footnote{More rigorously, $2\alpha$
  and  $2\beta$ are  H\"older-conjugated. Here we employ this  terminology for
  $\alpha$     and     $\beta$,     by     misuse    of     language.},     i.e.\
$\frac{1}{2\alpha}+\frac{1}{2\beta}=1$) as  they are based  on the Riesz--Thorin
theorem~\cite{HarLit52};  however  there  exist  few results  for  nonconjugated
indices. We summarize results available in the literature:
\begin{itemize}
\item  For  $\beta  \le   \frac{\alpha}{2  \alpha-1}$,  $N$-level  systems,  and
  $c=\frac{1}{\sqrt N}$  (complementary observables) : \  $\B_{\alpha,\beta; N} =
  \log N$ is the tight bound~\cite{Bia06,ZozPor08}.
\item  For   $\beta  >   \frac{\alpha}{2  \alpha-1}$,  $N$-level   systems,  and
  $c=\frac{1}{\sqrt N}$ : \ a known bound is $\overline{\B}_{\alpha,\beta; N} = 2
  \log  \left( \frac{2  \sqrt{N}}{1+\sqrt{N}}  \right)$~\cite[Eqs.~(7), (9)  and
  prop.~(d)]{MaaUff88}.  However,  this bound  is  not  tight.  Indeed, for  the
  particular case  $\alpha=\beta=2$ it was improved by  Luis~\cite{Lui07} to the
  value \ $2\log\left(\frac{2N}{N+1}\right)$.
\item For $\alpha=\beta=1$ (Shannon entropies), $N$-level systems, and arbitrary
  $c$  : \  $\overline{\B}_{1,1;  N} =  - 2  \log c$  was given  by Maassen  and
  Uffink~\cite{MaaUff88}.   This  bound has  been  improved  by  de Vicente  and
  Sanchez-Ruiz~\cite{VicSan08,BosPor11}  based on the  Landau--Pollak inequality
  linking   $\max_l  |\psi_l|^2$   and   $\max_k  |\widetilde{\psi}_k|^2$,   but
  restricted to the range $c\in[c^*,1]$ with $c^*\simeq 0.834$. These bounds are
  not     tight,      except     for     complementary      observables     (see
  also~\cite{Bia06,ZozPor08}) or for $N=2$~\cite{GhiMar03}.
\item For $\alpha =  \beta = \frac 12$, $N=2$ (qubits), and  arbitrary $c$ : the
  optimal       bound       $\B_{\frac        12,\frac       12;       2}       =
  \log\left[1+\sqrt{4c^2\left(1-c^2\right)}\right]$      was     obtained     by
  Rastegin~\cite{Ras12}.
\item For  $\alpha=\beta= 2$ (collision  entropies), $N=2$, and arbitrary  $c$ :
  the  tight  bound  was  found  by Bosyk  {\it  et  al.}~\cite{BosPor12},  with
  $\B_{2,2;2} = -2 \log \left( \frac{1+c^2}{2} \right)$.
%
\item For $\alpha=\beta\in(0, 2]$, $N=2$ and $c = 1/\sqrt{2}$, the optimal bound
  was   analyzed   in  the   context   of   the  Mach--Zehnder   interferometric
  setting~\cite{BosPor13}  with $\B_{\alpha,\alpha;2} =  \log 2$ if  $\alpha \le
  \alpha^\dag \approx  1.43$ and $\B_{\alpha,\alpha;2}  = \frac{2}{1-\alpha} \log
  \left[        \left(       \frac{1+1/\sqrt2}{2}\right)^2        +       \left(
      \frac{1-1/\sqrt2}{2}\right)^2 \right]$ otherwise.
\end{itemize}


\section{General R\'enyi entropic uncertainty relations for qubits} 
\label{Main:sec}

In  this  contribution  we  deal with the problem of 
generalizing  the  last  three 
developments
summarized in  the preceding section, i.e.\  for qubits ($N=2$)  and any overlap
$c$,  to  the case  of  arbitrary  R\'enyi-entropy  indices $(\alpha,\beta)$  to
measure uncertainty. We seek for the  minimum of the entropies' sum in this general
situation and  study, as well, those  states that saturate the  bound.  Our main
results are given by the following propositions:
\begin{proposition}
  Let us consider a pair of quantum observables $A$ and $B$ acting on a two-dimensional
  Hilbert space, and the  corresponding eigenbases $\{ |a_1\rangle , |a_2\rangle
  \}$ and  $\{ |b_1\rangle , |b_2\rangle  \}$. Consider a quantum  system in the
  qubit pure state $|\Psi\rangle$ described  by the projections $\psi = [ \psi_1
  \quad  \psi_2   ]^t$  or   $\widetilde{\psi}  =  [   \widetilde{\psi}_1  \quad
  \widetilde{\psi}_2 ]^t = T \psi$  on those bases respectively, where $T_{lk} =
  \langle b_l  | a_k  \rangle$ for  $k,l=1,2$. Then, for  any couple  of R\'enyi
  entropic indices $(\alpha,\beta) \in \Rset_+^2$,
  the following uncertainty relation holds:
\begin{equation}
H_\alpha(|\psi|^2) + H_\beta(|\widetilde{\psi}|^2) \ge \B_{\alpha,\beta; 2}(c)
\end{equation}
where the tight lower bound for the sum of R\'enyi entropies is obtained as
\begin{equation}
\B_{\alpha,\beta; 2}(c) = \min_{\theta \in [0 \: , \: \gamma]} \left( \frac{\log
\D_\alpha(\theta) }{1-\alpha} + \frac{\log \D_\beta(\gamma-\theta) }{1-\beta}
\right)
\label{Cab:eq}
\end{equation}
with
\begin{equation}
c \:\: \equiv \max_{k,l=1,2} | T_{lk}| \ \in \ \left[ \frac{1}{\sqrt{2}} \: , \:
1 \right] , \qquad \gamma \equiv \arccos c \qquad
\mbox{and} \qquad \D_\lambda(\theta) \equiv \left(\cos^2 \theta \right)^\lambda +
\left( \sin^2 \theta \right)^\lambda.
\label{c_D:eq}
\end{equation}
\label{UnceraintyQubitGeneral:prop1}
\end{proposition}

Furthermore, for any pair of two-dimensional observables we advance the minimizing solution.
\begin{proposition}
  Under the conditions of Proposition~\ref{UnceraintyQubitGeneral:prop1}, let us
  parameterize the matrix $T$ in the form~\cite{Dit03,Jar05}
\begin{equation}
T = \Phi(\vec{u}) \, V(\gamma_T) \, \Phi(\vec{v}) \qquad \mbox{where} \quad
\Phi(\vec{\cdot}) = \exp\big(\imath \, \diag(\vec{\cdot}) \big)  \quad \mbox{and} \quad
V(\gamma_T) = \left[ \begin{array}{cc}
\cos \gamma_T & \sin \gamma_T\vspace{2mm}\\
-\sin \gamma_T & \cos \gamma_T \end{array}\right] ,
\end{equation}
%
in terms  of $\gamma_T  \in \left[0 \:  , \:  \frac{\pi}{2} \right]$ 
and  the 2D
vectors  $\vec{u},\vec{v}$.
Denote   by   $\{  \theta_\opt^{(i)}   \}_{i   \in   \I}$   the  set of arguments  
that minimize the expression in Eq.~\eqref{Cab:eq}, where  $\I$ lists 
all the different possible solutions. 
Then the bound  is achieved (up to  a global phase factor) for  the qubits whose
projections onto the $A$-eigenbasis are
%
\begin{equation}
\psi_\opt^{(i,\varphi,n)} = e^{\imath \varphi} \Phi(-\vec{v}) \, \left[\begin{array}{c}
\cos \left( \varepsilon_T \theta_\opt^{(i)} + n \frac{\pi}{2} \right)\\[2.5mm]
\sin \left( \varepsilon_T \theta_\opt^{(i)} + n \frac{\pi}{2} \right)
\end{array}\right]
\quad \mbox{with}
\quad \varphi \in [0 \: , \: 2 \pi) ,
\quad \varepsilon_T = \sign \left( \frac{\pi}{4} - \gamma_T \right)
\quad \mbox{and} \quad n=0,1 
\end{equation}
\label{UnceraintyQubitGeneral:prop2}
\end{proposition}

We now concentrate on discussing  some derivations of our approach, and postpone
the proofs  of the  propositions to the  appendices.  To  start with, we  make a
connection with so-called Landau--Pollak uncertainty inequality~\cite{LanPol61}.
Although our proofs do not rely on this uncertainty relation, we can link {\em a
  posteriori}  both  results when  the  inequalities  are  saturated.  For  that
purpose, let us  introduce the probability vectors $P^A$  and $P^B$ respectively
issued    from    the    optimal    states    $\psi_\opt^{(i,\varphi,n)}$    and
$\widetilde{\psi}_\opt^{(i,\varphi,n)} = T \psi_\opt^{(i,\varphi,n)}$, namely
$$
P^A = \begin{bmatrix} \cos^2 \left(
    \varepsilon_T  \theta_\opt^{(i)} +  n  \frac{\pi}{2} \right)\vspace{2mm}  \\
  \sin^2    \left(   \varepsilon_T    \theta_\opt^{(i)}   +    n   \frac{\pi}{2}
  \right) \end{bmatrix} \qquad \mathrm{and} \qquad P^B = \begin{bmatrix} \cos^2 \left( \gamma_T -
    \varepsilon_T  \theta_\opt^{(i)} -  n  \frac{\pi}{2} \right)\vspace{2mm}  \\
  \sin^2  \left( \gamma_T  - \varepsilon_T  \theta_\opt^{(i)} -  n \frac{\pi}{2}
  \right) \end{bmatrix} .
$$
A  rapid
inspection of the different cases for $\gamma_T$ and $n$ allows us to obtain
\begin{equation}
\arccos \sqrt{\max_{k=1,2} P^A_k} + \arccos \sqrt{\max_{l=1,2} P^B_l} = \arccos c
\end{equation}
where $\arccos c =  \gamma=\min \left( \gamma_T , \frac{\pi}{2}-\gamma_T \right)
= \frac{\pi}{4}  - \left| \frac{\pi}{4} -\gamma_T  \right| \in \left[ 0  \: , \:
  \frac{\pi}{4}  \right]$.  This corresponds  precisely to  the equality  in the
Landau--Pollak  relation.   This relation  is  explicitly  used  by Maassen  and
Uffink~\cite{MaaUff88},  and by de  Vicente and  Sanchez-Ruiz~\cite{VicSan08} to
obtain their respective inequalities.

We mention also that in the general case of arbitrary R\'enyi-entropy indices to
measure uncertainty, the  bound $\B_{\alpha,\beta;2}$, Eq.~\eqref{Cab:eq}, has to
be  sought   numerically.   However,  for   indices  in  some  regions   of  the
$\alpha$--$\beta$  plane, we are  able to  obtain analytical  or semi-analytical
results. These are presented in the following corollaries.

\begin{corollary}
  In    the     context    of    Propositions~\ref{UnceraintyQubitGeneral:prop1}
  and~\ref{UnceraintyQubitGeneral:prop2},  if the  entropic indices  lie  into the
  square
  \begin{equation}
    (\alpha,\beta) \in \left[ 0 \: , \: \frac 12 \right]^2,
  \end{equation}
  there exists an analytical expression for the bound as
  \begin{equation}
    \B_{\alpha,\beta; 2}(c) = \frac{\log [ c^{2\lambda} + (1-c^2)^\lambda
      ]}{1-\lambda} \quad \mbox{where} \quad \lambda = \max(\alpha,\beta) .
  \end{equation}
  Moreover,  the wave-vectors  that  saturate the  inequality  correspond to:  \
  $\theta_\opt =  0 $ if $\alpha <  \beta$, $\theta_\opt = \gamma$  if $\alpha >
  \beta$, and both solutions if $\alpha = \beta$.

  \label{UnceraintyQubitSquare:prop}
\end{corollary}
First, one  can observe a transition in  terms of entropic indices  at $\alpha =
\beta$, since  only in  this situation both  angles lead to  wave-functions that
saturate         the          inequality.         We         notice         that
Corollary~\ref{UnceraintyQubitSquare:prop}  includes   some  of  the  situations
discussed at the end of Sec.~\ref{Position:sec} as particular cases.  On the one
hand,  when   $c$  is  fixed   to  $\frac{1}{\sqrt2}$,  the  optimal   bound  of
Refs.~\cite{Bia06,ZozPor08}  is recovered, and  if $\alpha  = \beta$  this bound
coincides with that  given in Ref.~\cite{BosPor13}. On the  other hand, when $c$
is  unrestricted and  if $\alpha  = \beta  = \frac  12$, one  recovers  the bound
obtained  in  Ref.~\cite{Ras12}.   We  stress  that these  results  have  proven
analytically  and   extended  its  scope  for   any  $c$  and   for  any  couple
$(\alpha,\beta)$ in the square $\left[ 0  \: , \: \frac 12 \right]^2$.

\begin{figure}[htbp]
\centerline{\includegraphics[height=4cm]{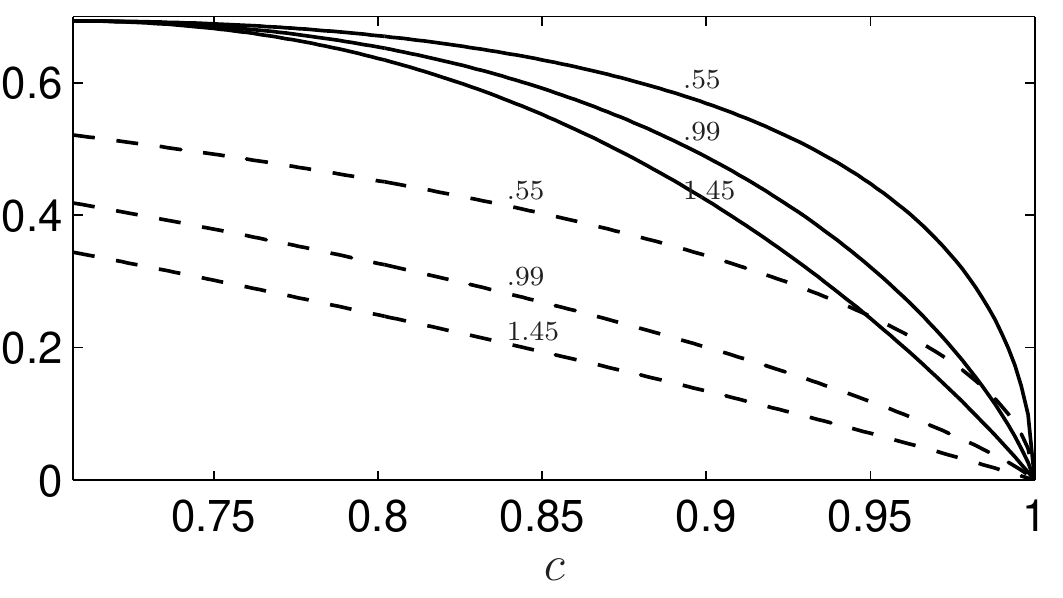}
\hspace{7.5mm} \includegraphics[height=4cm]{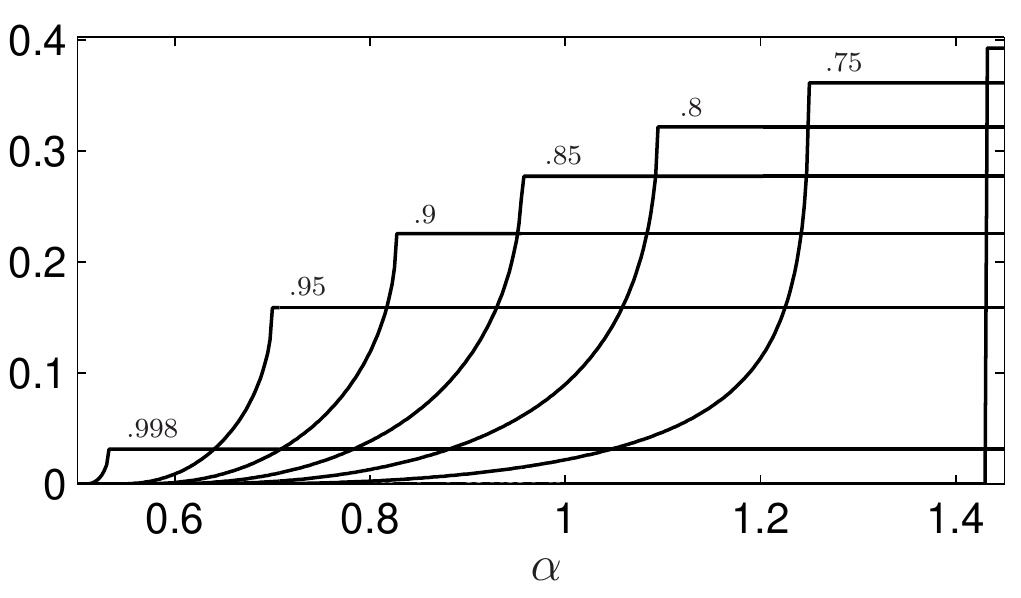}}
\caption{(a)   Functions  $\frac{\log   \left[  (c^2)^\alpha   +  (1-c^2)^\alpha
    \right]}{1-\alpha}$   (solid   lines)   and   $\frac{2\log   \left[   \left(
        \frac{1+c}{2}  \right)^\alpha   +  \left(  \frac{1-c}{2}  \right)^\alpha
    \right]}{1-\alpha}$ (dashed lines) vs $c$, for some values of $\alpha$.  (b)
  Optimal angle $\theta_\opt$ as a function of $\alpha > 0.5$ for some $c$.}
\label{Cotas13_ThetaOpt:fig}
\end{figure}

On the line $\beta = \alpha$, we obtain a semi-analytical result as follows:
\begin{corollary}
  In    the     context    of    Propositions~\ref{UnceraintyQubitGeneral:prop1}
  and~\ref{UnceraintyQubitGeneral:prop2},  if  the  entropic indices  are  equal
  ($\beta = \alpha$),
  the bound can be expressed as
  \begin{equation}
    \B_{\alpha,\alpha; 2}(c) = \left\{\begin{array}{lll} \frac{\log \left[
            (c^2)^\alpha + (1-c^2)^\alpha \right)}{1-\alpha} & \mbox{ if } & \alpha \in
        \left[ 0 \: , \: 0.5 \left( 1 - \delta_{c,\frac{1}{\sqrt{2}}} \right) + \alpha^\dag \, \delta_{c,\frac{1}{\sqrt{2}}}
        \right]\vspace{2.5mm}\\
        {\displaystyle \min_{\theta \in \left( 0 \: , \: \frac{\gamma}{2} \right]}}
        \frac{\log D_\alpha(\theta) + \log D_\alpha(\gamma-\theta)}{1-\alpha}
        & \mbox{ if } & \alpha \in \left[ 0.5 \left( 1 - \delta_{c,\frac{1}{\sqrt{2}}} \right) + \alpha^\dag \,
          \delta_{c,\frac{1}{\sqrt{2}}} \: , \: \alpha^\star(c) \right]\vspace{2.5mm}\\
        \frac{2 \, \log \left[ \left( \frac{1+c}{2} \right)^\alpha + \left( \frac{1-c}{2}
            \right)^\alpha \right]}{1-\alpha} & \mbox{ if } & \alpha \ge \alpha^\star(c)
      \end{array}\right.
    \label{CotaLine:eq}
  \end{equation}
  where  $\gamma = \arccos  c$, with  $\alpha^\dag \approx  1.43$ is  the unique
  solution  of  $\frac{2}{1-\alpha}   \log  \left[  \left(  \frac{2+\sqrt{2}}{4}
    \right)^\alpha +  \left( \frac{2-\sqrt{2}}{4} \right)^\alpha  \right] = \log
  2$ and  with $\alpha^\star(c)$ given by  both Fig.~\ref{AlphaEstrella:fig} and
  Table~\ref{AlphaEstrella:tab} and  where $\delta_{x,y}$ denotes  the Kronecker
  symbol.

\begin{figure}[htbp]
\begin{tabular}
{>{}m{.4\textwidth}
>{}m{.55\textwidth}
}
    \centerline{\includegraphics[width=7cm]{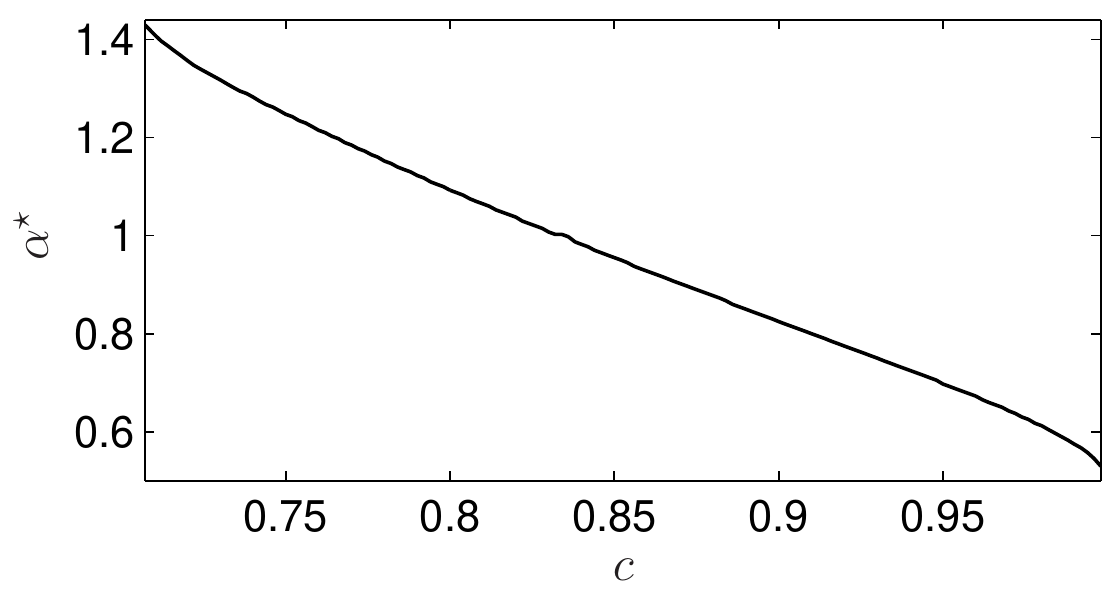}}
&
    \caption{Transition value  $\alpha^\star(c)$ for $c \in [1/\sqrt{2}  \: , \:
      1)$. For $c=1$ this value is  irrelevant since $\gamma = 0$ and both first
      and last expressions are equal to 0.\vspace{5mm}}
    \label{AlphaEstrella:fig}
\end{tabular}
\end{figure}

\begin{table}[htbp]
\begin{center}
\begin{tabular}{c||c|c|c|c|c|c|c|c|c|c|c|c|c|c|c}
$c$ & .71 & .73 & .75 & .77 & .79 & .81 & .83 & .85 & .87 & .89 & .91 & .93 & .95 & .97 & .99 \\
\hline
$\alpha^\star$ & 1.411 & 1.317 & 1.249 & 1.185 & 1.124 & 1.065 & 1.009 & .955 & .903 & .852 & .801 & .751 & .699 & .644 & .576
\end{tabular}
\end{center}
\caption{Transition values $\alpha^\star(c)$ for some $c \in (1/\sqrt{2} \: , \: 1)$.}
\label{AlphaEstrella:tab}
\end{table}

Moreover, the  bound is achieved  for $\theta_\opt^{(i)} = \frac{\gamma}{2}  + i
\left( \frac{\gamma}{2} - \theta_\opt \right)$ with $i \in \I = \{-1 , 1 \}$ and
$\theta_\opt = 0$ in the first regime, $\theta_\opt $ is the (unique, numerical)
solution  of  the  minimization  in   the  second  regime,  and  $\theta_\opt  =
\frac{\gamma}{2}$  in the last  regime (thus  the two  solutions reduce  to only
one).
    \label{UnceraintyQubitLine:prop}
\end{corollary}
From this corollary one can observe the following facts:
\begin{itemize}
\item  When $c  =  \frac{1}{\sqrt{2}}$, one  has  $\alpha^\dag =  \alpha^\star$.
  Thus, the  second expression in~\eqref{CotaLine:eq} reduces to  the first one,
  leading to  a transition in the  value of the bound  at $\alpha= \alpha^\dag$.
  This  can  also  be  seen  from  the  minimizers,  since  optimal  values  are
  $\theta_\opt = 0$,  or $\theta_\opt = \frac{\gamma}{2}$ or  both, depending on
  whether  $\alpha$ is  smaller than,  larger  than or  equal to  $\alpha^\dag$.
  These     observations    are    in     concordance    with     the    results
  in~\cite{BosPor13}.  Besides, the  value $\alpha^\dag$  in the  transition was
  already been observed implicitly in~\cite{Lui11}  as the index that vanish the
  second  derivative versus  $\theta$  of $\frac{\log  \D_\alpha(\theta) +  \log
    \D_\alpha(\gamma-\theta)}{1-\alpha}$ in $\theta = \frac{\gamma}{2}$.
\item  When  $\frac{1}{\sqrt{2}}  <   c  <  1$,  $\alpha^\star$  decreases  from
  $\alpha^\dag$ to  0.5. The  first and last  expressions in~\eqref{CotaLine:eq}
  reached and tend to  0 (see Fig.~\ref{Cotas13_ThetaOpt:fig}(a)). Moreover, the
  intermediate   expression,   the   optimal   angle   $\theta_\opt$   increases
  continuously        from       0       to        $\frac{\gamma}{2}$       (see
  Fig.~\ref{Cotas13_ThetaOpt:fig}(b)). In  this context, there  is no transition
  in the value  of the bound.  Moreover, $\alpha^\star$ is  not given in general
  as the index so that  the second derivative of $\frac{\log \D_\alpha(\theta) +
    \log  \D_\alpha(\gamma-\theta)}{1-\alpha}$  in  $\theta =  \frac{\gamma}{2}$
  vanish: the reasoning gave in~\cite{Lui11} does not hold this
  case. 
\item When $c=1$, one recovers then the trivial bound $\B_{\alpha,\alpha; 2}(1) =
  0$ and $\alpha^\star$ is irrelevant.
\end{itemize}

We   can   observe   that   some    situations   discussed   at   the   end   of
Sec.~\ref{Position:sec} are  include in this corollary as  particular cases.  On
the   one   hand,   for   $\alpha   \to   1$,   the   de   Vicente--Sanchez-Ruiz
bound~\cite{VicSan08} is recovered. Therefore,  it is optimal for qubit systems,
although  it was  calculated treating  separately $\psi$  and $\widetilde{\psi}$
without  taking into  account  the  relation between  them,  except through  the
Landau--Pollak inequality.   On the other hand,  for $\alpha =  2$, one recovers
the tight  bound obtained by Bosyk  \textit{et al.}~\cite{BosPor11}.  Therefore,
we  extend  previous results  along  all  the line  $\beta  =  \alpha$ giving  a
semi-analytical expression for the bound.

Finally,   note  that   using   the  fact   that   $H_\lambda$  decreases   with
$\lambda$~\cite{CovTho06,MaaUff88,HarLit52} one obtains the suboptimal result
\begin{corollary}
  In  the  context  of Proposition~\ref{UnceraintyQubitGeneral:prop1},  for  any
  couple $(\alpha,\beta) \in \Rset_+^2$, the entropies sum is lower bounded by
  \begin{equation}
    H_\alpha(|\psi|^2) + H_\beta(|\widetilde{\psi}|^2) \ge
    \overline{\B}_{\lambda,\lambda; 2}(c) \quad \mbox{with} \quad \lambda =
    \max(\alpha,\beta)
\end{equation}
where $\overline{\B}_{\lambda,\lambda; 2}(c)$ is given in Eq.~\eqref{CotaLine:eq}.
\label{UnceraintyQubitSubOpt:prop}
\end{corollary}
This bound  is clearly  suboptimal, as  it can be  see for  example in  the case
$(\alpha,\beta) \in \left[ 0 \: , \: \frac 12 \right]^2$.

\begin{figure}[htbp]
\begin{tabular}
{>{}m{.35\textwidth}
>{}m{.6\textwidth}}
\centerline{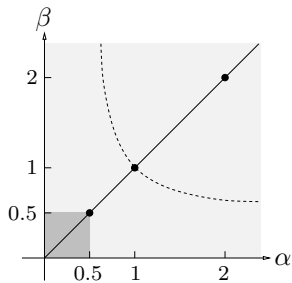}
&
\caption{Plane of entropic  indices. The bound is tight in  the square $\left[ 0
    \: ,  \: \frac 12  \right]^2$
  (dark grey region)
 and on  the line $\beta=\alpha$ (solid line).  Outside this region the
  bound is suboptimal (clear grey region)
  Note    that    previous    results    $(\alpha,\beta)=\left(\frac    12,\frac
    12\right)$,$(1,1)$ and  $(2,2)$ (dots) belong  to the first  region, whereas
  the conjugacy curve (dashed line) belongs to the second one.}
\label{regiones:fig}
\end{tabular}
\end{figure}

In   Fig.~\ref{regiones:fig},   we   represent   schematically  the   scope   of
Proposition~\ref{UnceraintyQubitSquare:prop},
Corollaries~\ref{UnceraintyQubitLine:prop}   and~\ref{UnceraintyQubitSubOpt:prop}
with  shadowed region,  solid line  and  dotted lines  in the  $\alpha$--$\beta$
plane.

Note  that starting from  $\alpha \to  \infty$ and  appealing to  the decreasing
property of the entropy versus the index, we recover the relation
\begin{equation}
H_\alpha(|\psi|^2) + H_\beta(|\widetilde{\psi}|^2) \ge - 2 \log \left(
\frac{1+c}{2} \right)
\end{equation}
which is precisely the bound  obtained by Deutsch~\cite{Deu83} in the context of
Shannon entropies,  or the  one given by  Maassen and  Uffink for any  couple of
indices   (before   being   refined   in  the   same   article)~\cite[Eq.~9   \&
prop.~(d)--(e)]{MaaUff88}.


\section{Discussion}
\label{Discussion:sec}

For  pure  states  of  qubit  systems,  we  obtain  the  most  general  entropic
formulation  of  the  uncertainty principle  in  terms  of  the sum  of  R\'enyi
entropies associated with any given pair of quantum observables, namely an inequality of
the    form     $H_\alpha(|\psi|^2)    +    H_\beta(|\widetilde{\psi}|^2)    \ge
\B_{\alpha,\beta; 2}(c)$ where  $c$ is the overlap of  the transformation between
the eigenspaces  of the  observables. Our derivation  focusses on  obtaining the
minimum of the entropies sum and we do not use Riesz--Thorin theorem in contrast
to many results in the literature.  In this way, we avoid the H\"older conjugacy
constraint on indices $\alpha$ and $\beta$  and our bound is tight and valid for
any couple of indices. Indeed, the bound obtained is universal in the sense that
it does not depend  on the state of the quantum system.  This is the main  result of the paper, given
in                                Propositions~\ref{UnceraintyQubitGeneral:prop1}
and~\ref{UnceraintyQubitGeneral:prop2}. Unfortunately,  we do not  always obtain
an analytical expression  for the bound.  However, we do  obtain in some domains
of  the   plane  $(\alpha,\beta)$  that   the  bound  takes  an   analytical  or
semi-analytical           expression.           In           effect,          in
Corollary~\ref{UnceraintyQubitSquare:prop}  we present an  analytical expression
for the tight bound in the square  $(\alpha,\beta) \in \left[ 0 \: , \: \frac 12
\right]^2$;
whereas  in Corollary~\ref{UnceraintyQubitLine:prop}  we show  a semi-analytical
expression for  the tight bound on  the line $\beta =  \alpha$.  Accordingly, we
recovered many  bounds derived  in the literature  for particular points  of the
plane  $(\alpha,\beta)$.  Moreover,  using  the nonincreasing  property of  the
R\'enyi entropy versus the entropic index, an analytical bound is obtained, this
last one being suboptimal (Corollary~\ref{UnceraintyQubitSubOpt:prop}). The same
propert allows also to recover the suboptimal bound primarily derived by Maassen
and Uffink.

For   mixed    states,   it    is   easy   to    extend   the    validity   of
Proposition~\ref{UnceraintyQubitGeneral:prop1}                                and
Corollaries~\ref{UnceraintyQubitSquare:prop}   and~\ref{UnceraintyQubitLine:prop}
within the domain $(\alpha,\beta)= [0,2]^2$  using the concavity property of R\'enyi entropy.
Indeed   for   $N$-level   systems,    if   one   has   a   universal   relation
$H_\alpha(|\psi|^2)          +         H_\beta(|\widetilde{\psi}|^2)         \ge
\overline{\B}_{\alpha,\beta;   N}$   satisfied   for   any   pure   state,   with
$(\alpha,\beta) \in [0 \: , \: \lambda_*(N)]^2$
then  one has  $H_\alpha \left(  \sum_m \mu_m  |\psi^{(m)}|^2 \right)  + H_\beta
\left( \sum_m  \mu_m |\widetilde{\psi}^{(m)}|^2 \right) \ge  \sum_m \mu_m \left(
  H_\alpha(|\psi^{(m)}|^2)  +  H_\beta(|\widetilde{\psi}^{(m)}|^2)  \right)  \ge
\sum_m \mu_m  \overline{\B}_{\alpha,\beta; N} =  \overline{\B}_{\alpha,\beta; N}$,
where $\sum_m \mu_m=1$. 
In  other words,  any  uncertainty formulation  for  pure states  in the  domain
$(\alpha,\beta) \in [0 \: , \:  \lambda_*(N)]^2$
remains
valid  for mixed states.   It remains  to be  studied the  way of  overcome this
constraint in the  domain of entropic indices due to  the concavity property and
the generalization of the results to $N$-level systems.


\paragraph{Acknowledgments:}  SZ  is  grateful  to  the  R\'egion  Rh\^one-Alpes
(France) for the grant that enabled  this work. GMB and MP acknowledge financial
support from CONICET and ANPCyT (Argentina).


\appendix


\section{Proof     of     Proposition~\ref{UnceraintyQubitGeneral:prop1}     and
  Corollaries~\ref{UnceraintyQubitSquare:prop}
  and~\ref{UnceraintyQubitLine:prop}}

\subsection{Simplification of the problem}
\label{appa1}

Since $\|\psi\|_2 = 1$, such a vector can be written under the form
\begin{equation}
\psi = \Phi({\vec{\varphi}}) \, s
\end{equation}
where $s \in \Sset^{(1)}$ the unit sphere on $\Rset^2$ (i.e.~the circle) and
where matrix $\Phi$ is diagonal and writes
\begin{equation}
\Phi(\vec{x}) = \exp(\imath \, \diag(\vec{x}))
\end{equation}

We parameterize the unitary matrix $T$  as the product of three unitary matrices
(see~\cite[Eqs.~(1)--(19)]{Jar05} or~\cite[Th.~1]{Dit03})
\begin{equation}
T = \Phi(\vec{u}) \, V(\gamma_T) \, \Phi(\vec{v}) \quad \mbox{ with }
V(\gamma_T) = \left[\begin{array}{cc}
\cos \gamma_T \, & \sin \gamma_T \\
&\\
- \sin \gamma_T & \cos \gamma_T
\end{array}\right], \quad \gamma_T \in \left[ 0 \: , \: \frac{\pi}{2} \right)
\label{T_Initial_Form:eq}
\end{equation}
(the  other possible  angles  can be  taken  into account  playing with  phases
$\vec{u}$ and $\vec{v}$).

Then,  from  the  relation~\eqref{Relation_psi_psitilde:eq} between  $\psi$  and
$\widetilde{\psi}$  and  from  the  form~\eqref{T_Initial_Form:eq} of  $T$,  one
obtains
\begin{equation}
\widetilde{\psi} = \Phi(\vec{u}) \, V(\gamma_T) \, \Phi(\vec{v} + \vec{\varphi})
\, s
\label{Psitilde_Initial:eq}
\end{equation}

Note  first  that the  overlap  does  not depend  on  the  phases,  namely $c  =
\max_{k,l} |T_{lk}| = \max_{k,l} |V_{lk}(\gamma_T)|$.  The goal is then to solve
the minimization problem
\begin{equation}
\B_{\alpha,\beta; 2} = \min_{\vec{\varphi},s} \left( H_\alpha(|\Phi(\vec{\varphi}) \, s|^2) +
H_\beta(|\Phi(\vec{u}) \, V(\gamma_T) \, \Phi(\vec{v}+\vec{\varphi}) \, s|^2)
\right)
\label{MiniInit:eq}
\end{equation}

The problem simplifies due to numerous invariances and symmetries.
\begin{itemize}
\item Invariance under  a phase shift applied to  the wavevector (multiplication
  by  a  matrix   $\Phi$):  $$H_\alpha(|\Phi(\vec{\varphi})  s|^2)  +  H_\beta(|
  \Phi(\vec{u}) V(\gamma_T)  \Phi(\vec{v}+\vec{\varphi})|^2) = H_\alpha(|s|^2) +
  H_\beta(|V(\gamma_T) \, \Phi(\vec{\varphi}+\vec{v}) s|^2)$$ $\vec{\varphi} \to
  \vec{v}  + \vec{\varphi}$  being  isomorphic, minimization~\eqref{MiniInit:eq}
  reduces to
  \begin{equation}
    \B_{\alpha,\beta; 2} = \min_{\vec{\varphi},s}\left( H_\alpha(|s|^2) +
      H_\beta(| V(\gamma_T) \, \Phi(\vec{\varphi}) \, s|^2) \right)
    \label{MiniPhase:eq}
  \end{equation}
  At this  step, one can  notice that the  bound depends only on  $\gamma_T$.
\item Additional invariance  under a permutation of the  components: the entropy
  does not depend on the order  of the components, thus, playing with the phases
  one  sees that  $$H_\alpha(|s|^2)  + H_\beta(|V(\gamma_T)  \Phi(\vec{\varphi})
  s|^2)  = H_\alpha(|s|^2)  + H_\beta(|V\left(\textstyle{\frac{\pi}{2}}-\gamma_T
  \right) \Phi(\vec{\varphi}-[\pi \quad 0]^t)  s|^2)$$ and thus the minimization
  problem~\eqref{MiniPhase:eq} reduces a step more,
  \begin{equation}
    \B_{\alpha,\beta; 2} = \min_{\vec{\varphi},s}\left( H_\alpha(|s|^2) + H_\beta(|
      V(\gamma) \, \Phi(\vec{\varphi}) \, s|^2) \right) \quad \mbox{where} \quad \gamma =
    \min \left( \gamma_T \: , \: \frac{\pi}{2} - \gamma_T \right) \in \left[ 0 \: ,
      \: \frac{\pi}{4} \right]
    \label{MiniPermut:eq}
  \end{equation}
  This    result   prove    now   that    the   bound    $\B_{\alpha,\beta;2}   =
  \B_{\alpha,\beta;2}(c)$ only depends on the overlap
\begin{equation}
c = \cos  \gamma =  \max_{k,l=1,2} |T_{lk}|
\end{equation}
\item Symmetries  and periodicities  on $s$: Note  that $s \in  \Sset^{(1)}$ can
  write in terms of angle as $s(\theta) = [\cos \theta \quad \sin \theta]^t$
  \begin{itemize}
  \item     $\pi$-periodicity:      clearly,     $$H_\alpha(|s(\theta)|^2)     +
    H_\beta(|V(\gamma)         \Phi(\vec{\varphi})        s(\theta)|^2)        =
    H_\alpha(|s(\theta+\pi)|^2)    +   H_\beta(|V(\gamma)   \Phi(\vec{\varphi})
    s(\theta+\pi)|^2)$$ so that one can restrict the search to $\theta\in \left[
      -\frac{\pi}{2} \: , \: \frac{\pi}{2} \right]$.
  \item $\frac{\pi}{2}$-symmetry:  playing with the permutations  and phases, it
    can   be   shown   that   $$H_\alpha(|s(\theta)|^2)   +   H_\beta(|V(\gamma)
    \Phi(\vec{\varphi})      s(\theta)|^2)      =      H_\alpha(|s(\theta      +
    \textstyle{\frac{\pi}{2}})|^2)  +  H_\beta(|V(\gamma) \Phi(J  \vec{\varphi})
    s(\theta +  \textstyle{\frac{\pi}{2}})|^2)$$ where $J =  \begin{bmatrix} 0 &
      1\\1 &  0\end{bmatrix}$, allowing  one to restrict  a little bit  more the
    interval $\theta \in \left[ - \frac{\pi}{4} \: , \: \frac{\pi}{4} \right]$.
  \item opposite  angle: we can finally  note that $s(-\theta)  = \Phi([0 \qquad
    \pi]^t)  s(\theta)$ so  that $$H_\alpha(|s(\theta)|^2)  + H_\beta(|V(\gamma)
    \Phi(\vec{\varphi})     s(\theta)|^2)    =     H_\alpha(|s(-\theta)|^2)    +
    H_\beta(|V(\gamma)  \Phi(\vec{\varphi}+[0   \quad  \pi]^t)  s(-\theta)|^2)$$
    allowing  to  retrict  a  step  more  to  $\theta  \in  \left[  0  \:  ,  \:
      \frac{\pi}{4} \right]$
  \end{itemize}
  From these symmetries, the problem restricts to
  \begin{equation}
    \B_{\alpha,\beta;2}(c) \:\: = \min_{\vec{\varphi},\theta \in \left[ 0 \: , \:
        \frac{\pi}{4} \right]} \left( H_\alpha(|s(\theta)|^2) + H_\beta(|V(\gamma)
      \Phi(\vec{\varphi}) s(\theta) |^2 \right)  \quad \mbox{where} \quad \gamma =
    \min \left( \gamma_T \: , \: \frac{\pi}{2} - \gamma_T \right)
  \label{MiniPiS4:eq}
  \end{equation}
\end{itemize}


\subsection{The trivial case $c = 1$}
\label{appa2}

In this  case, $\gamma =  0$ and $V(0)  = I$ the identity.  Clearly for $s  = [1
\quad  0]^t$  one  finds  both  $H_\alpha(|s|^2)  =  0$  and  $H_\beta(|V(0)  \,
\Phi(\vec{\varphi}) \, s|^2) = H_\beta(|s|^2) = 0$ and thus
\begin{equation}
\B_{\alpha,\beta; 2}(1) = 0
\end{equation}
%


\subsection{The general nontrivial case $c < 1$}
\label{appa3}

In this case, one has then $0  < \gamma \le \frac{\pi}{4}$.  Let us then proceed
in two  steps: (i)~fix $s$ (i.e.~$\theta$)  and minimize the  entropies sum over
phases  $\vec{\varphi}$;  (ii)~for  the   minimizing  phase,  that  depends  (in
principle) on $\theta$, $\vec{\varphi}(\theta)$, determine the value of $\theta$
that minimizes the entropies sum.


\subsubsection{Minimization over phase $\vec{\varphi}$}

Note   that   phases   $\vec{\varphi}$   play   a  role   only   on   the   term
$H_\beta(|V(\gamma)   \Phi(\vec{\varphi})   s(\theta)|^2)$.    Note  then   that
$H_\beta(|V(\gamma)  \, \Phi(\vec{\varphi})  s(\theta)|^2)$  is invariant  under
multiplication of the  argument by the scalar $\exp(\imath  \varphi')$, i.e.\ by
shifting  both components  of $\vec{\varphi}$  by the  same phase  shift.  Thus,
without loss of generality, one can consider $\vec{\varphi} = [- \varphi_2 \quad
\varphi_2]^t$ and thus
\begin{equation}
\left| V(\gamma) \, \Phi(\vec{\varphi}) \, s(\theta) \right|^2 =
\cos^2 \varphi_2 \left[\begin{array}{r}
\cos^2 (\gamma-\theta)\\
\sin^2 (\gamma-\theta)
\end{array}\right]
+ \sin^2 \varphi_2
\left[\begin{array}{r}
\cos^2(\gamma+\theta) \\
\sin^2(\gamma+\theta)
\end{array}\right]
\end{equation}

Clearly both mappings $\varphi_2 \mapsto - \varphi_2$ and $\varphi_2 \mapsto \pi
- \varphi_2$  leave the entropy unchanged  so that, without  loss of generality,
one can consider $\varphi_2 \in \left[ 0 \: , \: \frac{\pi}{2} \right]$ only and
the   solutions   are   modulo   $\pi$.    The  goal   is   then   to   minimize
$H_\beta(|V(\gamma)  \Phi(\vec{\varphi})   s(\theta)|^2)$  over  $\varphi_2  \in
\left[0 \: , \: \frac{\pi}{2} \right]$, that is equivalent to find the maximum
\begin{equation}
  \max_{\varphi_2 \in \left[0 \: , \: \frac{\pi}{2} \right]}
  \frac{\left(|V(\gamma) \Phi(\vec{\varphi}) s(\theta)|^2\right)^\beta}{\beta-1}
\end{equation}

\begin{figure}[htbp]
\begin{tabular}
{
>{}m{.4\textwidth}
>{}m{.55\textwidth}
}
\centerline{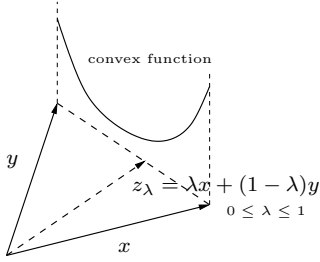}
&
\caption{\em  Vector  $z = \left|  V(\gamma)  \, \Phi(\vec{\varphi})  \,  s(\theta)
  \right|^2$   lives   in  the   segment   (the   convex)   that  link   vectors
  $x = [\cos^2(\gamma-\theta)        \quad       \sin^2(\gamma-\theta)]^t$       and
  $y = [\cos^2(\gamma+\theta) \quad  \sin^2(\gamma+\theta)]^t$. The typical behavior
  of a  convex function $\Rset^2  \to \Rset$ is  represented by the  solid line,
  illustrating that the maximum is attaind in the boundary of the convex set.}
\label{Convexity:fig}
\end{tabular}
\end{figure}

Note now  that $\left| V(\gamma) \, \Phi(\vec{\varphi})  \, s(\theta) \right|^2$
is   a  convex   combination  of   the  vectors   $[\cos^2(\gamma-\theta)  \quad
\sin^2(\gamma-\theta)]^t$        and        $[\cos^2(\gamma+\theta)        \quad
\sin^2(\gamma+\theta)]^t$  (see Fig.~\ref{Convexity:fig}).  Function  $x \mapsto
\frac{\|  x  \|_\beta^\beta}{\beta-1}$  being convex~\cite{HarLit52,Bul03},  the
maximum is attained  on the border of the convex set  defined by $\left\{ \left|
    V(\gamma) \, \Phi(\vec{\varphi})  \, s(\theta) \right|^2 \right\}_{\varphi_2
  \in \left[0  \: ,  \: \frac{\pi}{2} \right]}$  (see Fig.~\ref{Convexity:fig}),
namely
\begin{equation}
  \max_{\varphi_2 \in \left[0 \: , \: \frac{\pi}{2} \right]}
  \frac{\left(|V(\gamma) \Phi(\vec{\varphi}) s(\theta)|^2\right)^\beta}{\beta-1} =
  \max \left( \frac{\D_\beta(\gamma-\theta)}{\beta-1} \: , \:
    \frac{\D_\beta(\gamma+\theta)}{\beta-1}\right)
\end{equation}
where functions $\D_\lambda$ are defined by
\begin{equation}
\D_\lambda(\theta) = \left( \cos^2
\theta \right)^\lambda + \left( \sin^2 \theta \right)^\lambda
\label{D:eq}
\end{equation}
corresponding to $\varphi_2 = n \frac{\pi}{2}$, $n=0,1$.

\begin{figure}[htbp]
\begin{tabular}
{
>{}m{.4\textwidth}
>{}m{.55\textwidth}
}
\centerline{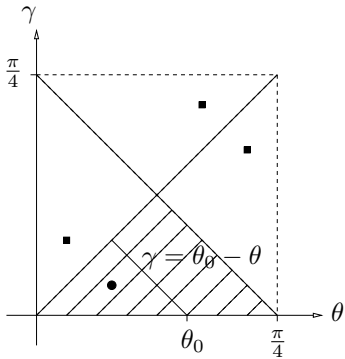}
&
\caption{\em  Illustration   of  the  symmetries  $\Delta(\theta , \gamma)  =
  \Delta(\gamma   ;  \theta) = \Delta(\pi/4-\theta , \pi/4-\gamma)  =
  \Delta(\theta  ; \gamma)$.   As an  illustration, the  point represented  by a
  circle  has  three  symmetric  (squares)  for  which  $\Delta$  has  the  same
  value.  Thus, (the  signum of)  function $\Delta$  has only  to be  studied on
  domain $D$  represented by zebras and given in Eq.~\eqref{TriangleDelta:eq}. The
  line in bold represents the  segment $\gamma = \theta_0-\theta$ in $D$, where
 function $\Delta$ is studied.}
\label{SymmetresDelta:fig}
\end{tabular}
\end{figure}

One can  go a step  further, comparing $\frac{\D_\beta(\gamma-\theta)}{\beta-1}$
and  $\frac{\D_\beta(\gamma+\theta)}{\beta-1}$.   To  this  end,   consider  the
difference
\begin{equation}
\Delta(\theta , \gamma) = \frac{\D_\beta(\gamma-\theta) - \D_\beta(\gamma+\theta)}{\beta-1}
\end{equation}
where $(\gamma,\theta) \in  \left[ 0 \: , \:  \frac{\pi}{4} \right]^2$. From the
obvious    symmetries   $\Delta(\theta,\gamma)    =    \Delta(\gamma,\theta)   =
\Delta(\frac{\pi}{4}-\theta,\frac{\pi}{4}-\gamma)$,  one can restrict  the study
to the triangle
\begin{equation}
D = \{(\theta,\gamma) \in [0 \: , \: \pi/4]^2, 0 \le \theta \le \min(\gamma ,
\pi/4-\gamma) \}
\label{TriangleDelta:eq}
\end{equation}
as depicted in Fig.~\ref{SymmetresDelta:fig}.  Note first that $\Delta(\theta,0)
= 0$,  then let us  study the  variations on the  segments $\gamma =  \theta_0 -
\theta$ as pictured in Fig.~\ref{SymmetresDelta:fig}.

A simple derivation leads to
\begin{equation}
\frac{\partial \Delta(\theta,\theta_0-\theta)}{\partial \theta} = \frac{2 \beta
\sin(4 \theta-2 \theta_0) \left( \left(\sin^2(2 \theta-\theta_0) \right)^\beta -
\left(\cos^2(2 \theta-\theta_0) \right)^\beta \right)}{\beta-1}
\end{equation}
Since $0  \le \theta_0/2 \le  \theta \le  \theta_0 \le \pi/4$,  on has $0  \le 2
\theta-\theta_0 \le \pi/4$,  and thus on the one  hand $\sin(4 \theta-2\theta_0)
\ge  0$   while  on   the  other  hand   $\sin(2  \theta-\theta_0)   \le  \cos(2
\theta-\theta_0)$.  This proves  that the derivative is positive,  and thus that
$\Delta(\theta,\theta_0-\theta)$   increases  with   $\theta$.    Together  with
$\Delta(\theta_0,0) =  0$ and the  symmetries, we have  proved then that  in the
whole   square   $\left[0   \:   ,   \:   \frac{\pi}{4}   \right]^2$   one   has
$\Delta(\theta,\gamma)  \ge  0$,  i.e.\  $\frac{\D(\gamma-\theta)}{\beta-1}  \ge
\frac{\D(\gamma+\theta)}{\beta-1}$.  This finishes to prove that
\begin{equation}
  \min_{\varphi_2 \in \left[0 \: , \: \frac{\pi}{2} \right]} H_\beta(|V(\gamma)
  \Phi(\vec{\varphi}) s(\theta)|^2) = \frac{\log \D_\beta(\gamma-\theta)}{1-\beta}
\end{equation}
obtained for
\begin{equation}
\varphi_{2,\mathrm{opt}} = 0
\end{equation}

Note that due to the invariance of the entropies sum under the multiplication of
the wavevector by a scalar $\exp(\imath \varphi')$, in some sense there exists a
``unique''  phase  $\vec{\varphi}$ minimizing  the  entropies  sum  when $s$  is
fixed. Moreover, ``this'' phase does not depend on $s$.


\subsubsection{Minimization over the angle $\theta$}

Before   specializing   the   problem   in   different  parts   of   the   plane
$(\alpha,\beta)$, one can simplify one  step more the interval in $\theta$ where
the minimum of  the entropies sum has to be sought.  From the preceding section,
minimization~\eqref{MiniPiS4:eq} reduces to
\begin{equation}
\B_{\alpha,\beta;2}(c) = \min_{\theta \in \left[0 \: , \: \frac{\pi}{4} \right]}
\left( \frac{\log \D_\alpha(\theta)}{1-\alpha} + \frac{\log
\D_\beta(\gamma-\theta)}{1-\beta} \right) \quad \mbox{with} \quad
D_\lambda(\theta) = \left( \cos^2 \theta \right)^\lambda + \left( \sin^2 \theta
\right)^\lambda
\end{equation}
Deriving the functional to minimize gives
\begin{equation}
\frac{\partial}{\partial \theta} \left( \frac{\log \D_\alpha(\theta)
}{1-\alpha} + \frac{\log \D_\beta(\gamma-\theta)}{1-\beta}
\right) = \frac{\D_\alpha'(\theta)}{(1-\alpha) \, \D_\alpha(\theta)} -
\frac{\D_\beta'(\gamma-\theta)}{(1-\beta) \, \D_\beta(\gamma-\theta)}
\label{FirstDerivativeEntropySum:eq}
\end{equation}
where the derivative in $\theta$ of functions $\D_\lambda$ writes
\begin{equation}
\D_\lambda'(\theta) \: \equiv \: \frac{\partial \D_\lambda}{\partial \theta}
= \lambda \sin (2 \theta) \, \left( \left( \sin^2 \theta
\right)^{\lambda-1} - \left( \cos^2 \theta \right)^{\lambda-1} \right)
\label{DerivativeD:eq}
\end{equation}
Since $\theta \in [0 \: , \: \pi/4]$, one has both $\sin ( 2 \theta ) \ge 0$ and
$\sin^2 \theta \le \cos^2 \theta$ and  thus the first fraction of the right-hand
side   (rhs)  of~\eqref{FirstDerivativeEntropySum:eq}  is   positive.  Moreover,
$\theta-\gamma \in  [-\pi/4 \: , \: \pi/4]$  and thus by the  same reasoning the
second fraction of the  rhs of~\eqref{FirstDerivativeEntropySum:eq} has the same
signum as $\sin ( 2 (\gamma - \theta)  )$. Thus, for $\theta \in (\gamma \: , \:
\pi/4]$  the  entropies sum  is  increasing.   Necessarily  the minimum  of  the
entropies sum is given by $0 \le \theta \le \gamma$, reducing the interval where
$\theta$ has to be sought, i.e.
\begin{equation}
\B_{\alpha,\beta; 2}(c) = \min_{\theta \in \left[ 0 \: , \: \gamma \right]} \left(
\frac{\log \D_\alpha(\theta)}{1-\alpha} + \frac{\log 
\D_\beta(\gamma-\theta)}{1-\beta} \right)
\label{SumEntropyThetaReduced:eq}
\end{equation}

At  this  step, the  minimum  of~\eqref{SumEntropyThetaReduced:eq}  can only  be
sought numerically, leading to Proposition~\ref{UnceraintyQubitGeneral:prop1}.

One can go a step further in special cases as we will see now.


\subsection{Special context $(\alpha,\beta) \in \left[ 0 \: , \: \frac 12\right]^2$.}

Let us start from Eq.~\eqref{FirstDerivativeEntropySum:eq}, that gives the
second derivative of the function to be minimize under the form
\begin{equation}
\frac{\partial^2} {\partial \theta^2} \left( \frac{\log \D_\alpha(\theta)
}{1-\alpha} + \frac{\log \D_\beta(\gamma-\theta)}{1-\beta}
\right) = \frac{\K_\alpha(\theta)}{\left( \D_\alpha(\theta) \right)^2} +
\frac{\K_\beta(\gamma-\theta)}{\left( \D_\beta(\gamma-\theta) \right)^2}
\label{SecondDerivativeEntropySum:eq}
\end{equation}
where functions $\K_\lambda = \left( \D_\lambda'' \, \D_\lambda -
\D_\lambda'^{\, 2} \right)/(1-\lambda)$ writes
\begin{equation}
\K_\lambda(\theta) = \frac{2 \, \lambda}{1-\lambda} \left( (2 \lambda-1) \left(
\frac{\sin^2 (2 \theta)}{4} \right)^{\lambda-1} - \left( \cos^2 \theta
\right)^{2 \lambda-1} \! - \left( \sin^2 \theta \right)^{2 \lambda-1} \right)
\label{K:eq}
\end{equation}
(one uses alternatively the fact that $\cos^2\theta - \sin^2\theta = 2
\cos^2\theta - 1 = 1 - 2 \sin^2\theta$ and the identities $\cos^2\theta
\sin^2\theta = \cos^2\theta - \cos^4\theta = \sin^2\theta -
\sin^4\theta$).

For  $\lambda \le  \frac 12$,  since one  has  then also  $\lambda <  1$ we  can
immediately conclude that the second derivative in $\theta$ of the entropies sum
is strictly negative, so that  function $\theta \mapsto \left( \frac{\log 
      \D_\alpha(\theta)}{1-\alpha}     +      \frac{\log 
      \D_\beta(\gamma-\theta)}{1-\beta}  \right)$ is concave  on $\theta
\in [0 \: , \gamma ]$  (the opposite function is convex) \cite{Bul03}. Thus, the
minimum (the maximum of the oposite) is  attain in the border of the convex line
$[0 \: , \gamma ]$, i.e. either for $\theta = 0$, or for $\theta = \gamma$.
%
%
It remains then to compare the values of the function  at these extremal points,
i.e.\     from~\eqref{SumEntropyThetaReduced:eq}--\eqref{D:eq}     to    compare
$\frac{\log         \D_\alpha(\gamma)}{1-\alpha}$         and        $\frac{\log
  \D_\beta(\gamma)}{1-\beta}$  with $D_\lambda$ given in Eq.~\eqref{c_D:eq}. Since
the   R\'enyi   entropy   $H_\lambda$    is   a   decreasing   function   versus
$\lambda$~\cite{CovTho06,MaaUff88,HarLit52},       together       with       the
expression~\eqref{D:eq}  of  $D_\lambda$  and  $c  = \cos  \gamma$  one  obtains
Corollary~\ref{UnceraintyQubitSquare:prop}.



\subsection{Semi-analytical results when $\alpha = \beta$.}

In order to simplify the notation, let us denote the function to minimize as
\begin{equation}
\F_\alpha(\theta) = \frac{\log \D_\alpha(\theta) + \log 
\D_\alpha(\gamma-\theta)}{1-\alpha}
\end{equation}
with  $\D_\alpha$  defined in Eq.~\eqref{D:eq},  so  that  the  problem is  to
minimize $\F_\alpha$ over $\theta \in [0 \: , \: \gamma]$. One can go a step further
observing the  trivial symmetry $\F_\alpha(\theta)  = \F_\alpha(\gamma-\theta)$,
so the the problem reduces to
\begin{equation}
\B_{\alpha,\alpha; 2}(c) = \min_{\theta \in \left[ 0 \: , \: \frac{\gamma}{2}
\right]} \F_\alpha(\theta)
\end{equation}

Since the case  $(\alpha,\beta) \in [0 \: , \: \frac  12]^2$ is already treated,
one concentrates  here in the context $\alpha  = \beta > \frac  12$. Recall also
that we  exclude here the case  $\alpha = 1$: it  will be recover  by taking the
limit $\alpha \to 1$.

Deriving $\F_\alpha$ versus $\theta$ leads to
\begin{equation}
\F_\alpha'(\theta) = \frac{\partial}{\partial \theta} \F_\alpha(\theta) =
\frac{1}{1-\alpha} \left( \frac{\D_\alpha'(\theta)}{\D_\alpha(\theta)} -
\frac{\D_\alpha'(\gamma-\theta)}{\D_\alpha(\gamma-\theta)} \right)
\end{equation}
where    $\D'_\alpha$   was    already    explicited in Eq.~\eqref{DerivativeD:eq}.
$\F_\alpha'$  clearly vanish  when  $\theta =  \frac{\gamma}{2}$  which gives  a
possible solution. The question now is  to determine if the extremum is a global
minimum or not.


\subsubsection{When $c = 1/\sqrt{2}$}

In this case, $\gamma = \frac{\pi}{4}$ and it is easy to see that $\D_\alpha'(0)
=  \D_\alpha'(\gamma)  =  0$.  Thus,  $\theta  = 0$  gives  another  solution  to
$\F_\alpha'(\theta) = 0$.

We observe then the following behaviours, already noticed in~\cite{BosPor13} : 
\begin{itemize}
\item $\theta  = 0$  is the  unique solution for  $\alpha <  \alpha^\dag \approx
  1.430$, leading to the bound  $\frac{\log \left[ c^{2 \alpha} + (1-c^2)^\alpha
    \right]}{1-\alpha} = \log 2$.
\item When  $\alpha >  \alpha^\dag$ the  unique solution is  given by  $\theta =
  \frac{\gamma}{2}$,  leading  to  the  bound  $\frac{2 \,  \log  \left[  \left(
        \frac{1+c}{2}  \right)^\alpha   +  \left(  \frac{1-c}{2}  \right)^\alpha
    \right]}{1-\alpha}  =  \frac{2 \,  \log  \left[ \left(  \frac{2+\sqrt{2}}{4}
      \right)^\alpha     +     \left(    \frac{2-\sqrt{2}}{4}     \right)^\alpha
    \right]}{1-\alpha}$.
\item When $\alpha  = \alpha^\dag$, there is a phase  transition: both $\theta =
  0$  and $\theta =  \frac{\gamma}{2}$ give  the minimum  bound $\log  2$, while
  $\alpha^\dag$  is  the unique  solution  of  $\frac{2  \, \log  \left[  \left(
        \frac{2+\sqrt{2}}{4}   \right)^\alpha   +  \left(   \frac{2-\sqrt{2}}{4}
      \right)^\alpha  \right]}{1-\alpha} =  \log 2$.  Note that  the  unicity is
  insured from  the the fact  that the entropy  is a decreasing function  of the
  index~\cite{CovTho06,MaaUff88,HarLit52}.
\end{itemize}


\subsubsection{When $c \in (1/\sqrt{2} \: , \: 1)$}

In this case, one still has $\D_\alpha'(0) = 0$ but clearly for any $\theta \in
\left(  0 \:  , \: \frac{\pi}{4}  \right)$,  $\frac{\D_\alpha'(\theta)}{1-\alpha} >
0$.  In other  words, $\F_\alpha'(0)  < 0$  so  that $\theta  = 0$  cannot be  a
solution to  $\F_\alpha'(\theta) =  0$: all the  possible solutions are  then in
$\left( 0 \: , \: \frac{\gamma}{2} \right]$.

We observe numerically the following behavior: For any fixed $c$, there exist a
$\alpha^\star(c)$ so that
\begin{itemize}
\item  for  $\alpha  \in (0.5  \:  ,  \:  \alpha^\star(c))$, the  optimal  angle
  $\theta_\opt(\alpha)$ is in  $ \left( 0 \: ,  \: \frac{\gamma}{2} \right)$ and
  can only be numerically seek. The bound is then numerically expressed as well.
  Moreover,   $\theta_\opt(\alpha)$   increases   continuously   from   $0$   to
  $\frac{\gamma}{2}$.
\item For $\alpha  \ge \alpha^\star(c)$ the only minimum is  given for $\theta =
  \frac{\gamma}{2}$,  leading  to  the  bound  $\frac{2 \,  \log  \left[  \left(
        \frac{1+c}{2}  \right)^\alpha   +  \left(  \frac{1-c}{2}  \right)^\alpha
    \right]}{1-\alpha}$; thus, there is no transition phase in this context.
\end{itemize}



\section{Expression of the minimizers}


\subsection{
Proof of Proposition~\ref{UnceraintyQubitGeneral:prop2}}

Let us first recall that the entropies sum is insensitive to the multiplication of the wavector by a scalar $e^{\imath \varphi}$, 
and thus 
from a  minimizer $\psi_0$  we will  obtain families of  minimizers of  the form
$\{e^{\imath \varphi} \psi_0 \}_{\varphi \in [0 \: , \: 2 \pi)}$.

Recall that any unitary matrix $T$ can be parameterized under the form
%
\begin{equation}
T = \Phi(\vec{u}) \, V(\gamma_T) \, \Phi(\vec{v}) \quad \mbox{with} \quad
V(\gamma_T) = \left[ \begin{array}{cc} \cos \gamma_T & \sin
\gamma_T\vspace{2mm}\\-\sin \gamma_T & \cos \gamma_T \end{array}\right], \quad
\Phi(\vec{x}) = \exp(\imath \, \diag(\vec{x})), \quad \gamma_T \in \left[0 \: , \:
\frac{\pi}{2} \right]
\label{ParamT:eq}
\end{equation}
(see~\cite[Eqs.~(1)--(19)]{Jar05} or~\cite[Th.~1]{Dit03}).  Let us denote by $\{
\theta_\opt^{(i)} \}_{i \in \I}$, the ensemble of the minimizers angles in $[0 \:
, \:  \gamma]$ of Eq.~\eqref{Cab:eq}--Eq.~\eqref{SumEntropyThetaReduced:eq} where
$\I$  indices   each  solution.

\begin{itemize}
\item Case  $\gamma_T = \gamma  \in \left[0 \:  , \: \frac{\pi}{4}  \right]$. We
  have that $T \psi =  \Phi(\vec{u}) V(\gamma) \Phi(\vec{v}) \psi$ so that
  an  ensemble  of  minimizers   takes  the  form  $\left\{  e^{\imath  \varphi}
    \Phi(-\vec{v})   s(\theta_\opt^{(i)})  \right\}_{i  \in   \I,  \varphi\in
    \left[0 \: , \: 2 \pi \right)}$.
  \begin{itemize}
  \item Symmetry $\theta  \to - \theta$: From the study  of the preceding section,
    it is  immediate that\newline $\left\{ e^{\imath  \varphi} \Phi(-\vec{v}+[0 \quad
      \pi]^t) s(-\theta_\opt^{(i)}) \right\}_{i \in  \I, \varphi\in \left[0 \: ,
        \: 2 \pi \right)}$ is also a  family of minimizers. It turns out that it
    is the same family, than that obtained from the $\theta_\opt^{(i)}$.
  \item Symmetry $\theta \to \theta + \frac{\pi}{2}$: From the preceding section,
    from  angles $\theta_\opt^{(i)}$ one  obtains now  the family  of minimizers
    $\left\{     e^{\imath      \varphi}     \Phi(-    \vec{v})     s\left(
        \theta_\opt^{(i)}+\frac{\pi}{2}  \right) \right\}_{i \in  \I, \varphi\in
      \left[0 \: , \: 2 \pi \right)}$.
  \item  Symmetry $\theta  \to  \theta +  \pi$:  Clearly, one  obtains the  same
    families (starting respectively from  the $\theta_\opt^{(i)}$ and from the the
    $\theta_\opt^{(i)}+\frac{\pi}{2}$).
  \end{itemize}
  In  a   conclusion,  for   $\gamma_T  =  \gamma$   the  minimizers   take  the
  form     $$\psi_\opt^{(i,\varphi,n)}    =     e^{\imath     \varphi}    \Phi(-
  \vec{v})    \begin{bmatrix}    \cos    \left(   \theta_\opt^{(i)}    +    n
      \textstyle{\frac{\pi}{2}}    \right)    \vspace{2.5mm}\\    \sin    \left(
      \theta_\opt^{(i)}  +  n  \textstyle{\frac{\pi}{2}}  \right)  \end{bmatrix}
  \quad \mbox{with}  \quad i \in  \I, \quad  n=0,1  \quad \mbox{and}
  \quad \varphi \in [0 \: , \: 2 \pi) $$
\item Case  $\gamma_T = \frac{\pi}{2}-\gamma  \in \left[0 \: ,  \: \frac{\pi}{4}
  \right]$.    We   have   now   $T   \psi  =   \Phi(\vec{u})   J   V(\gamma)
  \Phi(\vec{v}+[0 \quad \pi]^t) \psi$ so that an ensemble of minimizers takes
  the  form  $\left\{   e^{\imath  \varphi}  \Phi(-\vec{v}-[0  \quad  \pi]^t)
    s(\theta_\opt^{(i)}) \right\}_{i  \in \I, \varphi\in  \left[0 \: , \:  2 \pi
    \right)}    =    \left\{    e^{\imath   \varphi}    \Phi(-\vec{v})    s(-
    \theta_\opt^{(i)})  \right\}_{i \in  \I, \varphi\in  \left[0 \:  , \:  2 \pi
    \right)}$.
 Then,  following the same steps  than before, one  obtains in this
  case the family of minimizers $$\psi_\opt^{(i,\varphi,n)} = e^{\imath \varphi}
  \Phi(-  \vec{v})  \begin{bmatrix}  \cos  \left(  -  \theta_\opt^{(i)}  +  n
      \textstyle{\frac{\pi}{2}}   \right)   \vspace{2.5mm}\\   \sin   \left(   -
      \theta_\opt^{(i)}  +  n  \textstyle{\frac{\pi}{2}}  \right)  \end{bmatrix}
  \quad \mbox{with}  \quad i \in  \I, \quad  n=0,1  \quad \mbox{and}
  \quad \varphi \in [0 \: , \: 2 \pi) $$
\end{itemize}
One  can  unify  both  cases  by   noting  than  the  signum  before  the  angle
$\theta_\opt^{(i)}$ is  nothing moe  than $\sign\left( \frac{\pi}{4}  - \gamma_T
\right)$,       leading       to       the       expression       given       in
Proposition~\ref{UnceraintyQubitGeneral:prop2}.


\subsection{A step more in the case $\alpha = \beta$.}

\begin{figure}[htbp]
\centerline{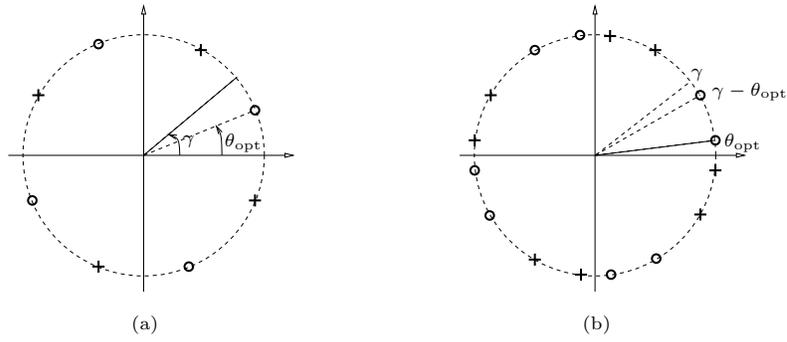}
\caption{(a) Angles issued from the optimal one in the case where $\gamma_T \in
\left[ 0 \: , \: \frac{\pi}{4} \right]$ (circle) and where $\gamma_T \in
\left[ \frac{\pi}{4} \: , \: \frac{\pi}{2} \right]$ (crosses) in the general
case. For sake of readness, in this illustration we assume that $\theta_\opt$
is unique. (b) Same as in~(a), when $\alpha = \beta$, taking into account the
symmetries in this special case.}
\label{AnglesOpt:fig}
\end{figure}

One has  seen numerically the existence  of a unique  optimal angle $\theta_\opt
\in  \left[ 0  \: ,  \: \frac{\gamma}{2}  \right]$ (with  $\gamma =  \arccos c$)
leading to the  minimal bound of the entropies sum. Moreover,  we have seen that
the entropies sum is invariant under the transformation $\theta \to \gamma - 
\theta$.     This    leads   to    the    possible    angles   represented    in
Fig.~\ref{AnglesOpt:fig}(b),  respectively for $\gamma_T  \in \left[  0 \:  , \:
  \frac{\pi}{4} \right]$  (circles) and $\gamma_T \in \left[  \frac{\pi}{4} \: ,
  \: \frac{\pi}{2}  \right]$ (crosses).  As a  conclusion, $$\theta_\opt^{(i)} =
\frac{\gamma}{2}  +  i  \left(  \frac{\gamma}{2}  -  \theta_\opt  \right)  \quad
\mbox{with} \quad i \in  \I = \{-1 , 1 \},$$ leading  to the minimizers given in
corolary~\ref{UnceraintyQubitLine:prop}.

\bibliography{Uncertainty_qubit9}
\bibliographystyle{unsrt}

\end{document}